\documentclass[conference]{IEEEtran}
\usepackage{cite}
\usepackage{amsmath,amssymb,amsfonts}
\usepackage{algorithmic}
\usepackage{graphicx}
\usepackage{textcomp}
\usepackage{xcolor}
\usepackage{tikz}
\usetikzlibrary{positioning}
\usepackage{subcaption}
\usepackage{xcolor}
\usepackage{mathtools, nccmath}
\usepackage{comment}

\DeclarePairedDelimiter{\nint}\lfloor\rceil

\def\BibTeX{{\rm B\kern-.05em{\sc i\kern-.025em b}\kern-.08em
    T\kern-.1667em\lower.7ex\hbox{E}\kern-.125emX}}
\begin{document}

\title{Optimization of Actuarial Neural Networks with Response Surface Methodology\\}

\author{\IEEEauthorblockN{Author: Belguutei Ariuntugs}
\IEEEauthorblockA{\textit{Department of Mathematics} \\
\textit{Tennessee Technological University.}\\
Cookeville, 38501, TN, USA \\
bariuntug42@tntech.edu}
\and
\IEEEauthorblockN{Co-author: Kehelwala Dewage Gayan Maduranga}
\IEEEauthorblockA{\textit{Department of Mathematics} \\
\textit{Tennessee Technological University}\\
Cookeville, 38501, TN, USA \\
gmaduranga@tntech.edu}
}

\maketitle

\begin{abstract}
In the current data-driven landscape, machine learning (ML) is pivotal, especially in actuarial science, where precision is paramount. Neural networks are powerful tools for predictive modeling, improving risk assessment, pricing strategies, and decision making within the insurance and financial sectors. These algorithms require the optimization of user-defined hyperparameters, such as learning rates, number of layers, and activation functions, to ensure efficient use of resources as the complexity and scale of ML algorithms increase. This study focuses on combined actuarial neural networks (CANN), selected for its relevance in actuarial applications and built-in safety features for tasks such as mortality forecasting and pricing.

We employ a factorial arrangement design of experiments for sampling hyperparameters, followed by response surface methodology (RSM) to fit a quadratic surface over the hyperparameter space, identifying optimal or near-optimal settings. Our proposed method is particularly well-suited for optimizing performance when resources are constrained. Unlike traditional methods such as grid search, which test multiple hyperparameter combinations sequentially and can require many runs, RSM offers a structured approach to experiment design that effectively explores the response surface and captures potential curvature.

Results show that our method not only predicts the performance of the CANN algorithm accurately, but also identifies critical hyperparameters. We investigated two cases of optimization: one in which every hyperparameter is considered significant and another in which statistically insignificant ones are dropped after the first phase of the experiment. Ultimately, it was revealed that when every hyperparameter is tuned, we reached the minimum loss in 288 runs with a 0.245823 out-of-sample Poisson deviance loss. When two statistically insignificant hyperparameters were dropped, we reached near-minimum loss in 188 runs with a 0.245976 out-of-sample Poisson deviance loss. Thus, we established that we could achieve reasonable performance by omitting less significant terms, thereby reducing computational costs without substantially sacrificing accuracy.
\end{abstract}

\begin{IEEEkeywords}
Actuarial Science, Design of Experiments, Hyperparameter Optimization, Response Surface Methodology, Neural Network\\

This work was presented at the Actuarial Research Conference (ARC) 2024.\footnote{Research abstract submitted and presented at ARC 2024. More details can be found at \url{https://sites.google.com/view/arc2024/home}.}

\end{IEEEkeywords}

\section{Introduction}
In the realm of actuarial science, where precision and accuracy are paramount, the integration of advanced computational techniques has become increasingly indispensable. One such technique, neural networks, has emerged as a powerful tool for predictive modeling, offering the potential to enhance risk assessment, pricing strategies, and decision-making processes within the insurance and financial sectors \cite{Richman2022}.

Neural networks can improve performance and efficiency in various aspects, such as handling large amounts of data and recognizing patterns to perform tasks like prediction and classification. Prediction involves estimating some quantitative outcome of a future observation based on the values of other attributes, while classification entails assigning labels to unlabeled data instances. Deep neural network predictions have been shown to outperform traditional actuarial techniques in pricing \cite{Schelldorfer2019}, claims reserving \cite{Gabrielli2018}, and mortality forecasting \cite{Hainaut2018}. We have chosen Combined Actuarial Neural Networks (CANN) for its relevance in actuarial applications and built-in safety features for tasks such as mortality forecasting and pricing. 

The efficacy of neural networks crucially depends on the selection of their hyperparameters. Hyperparameters are a set of parameters that must be configured before the training process by the modeler. They govern the network's architecture and learning dynamics, and even a small change in hyperparameters can greatly affect the model's performance. Each hyperparameter has a wide range of possible values that a modeler must choose from, yet the optimal set of hyperparameters remains unknown. This requires an efficient search strategy that is both economical and fast. Consequently, hyperparameter optimization becomes a critical endeavor to unlock the full potential of neural networks in actuarial applications.

In this paper, we propose optimizing the hyperparameters of CANN, developed by Wüthrich and Merz \cite{Wuthrich2019}, using the Response Surface Methodology (RSM) within the Design of Experiments (DOE) framework. RSM allows us to create a quadratic response surface across the hyperparameter space, enabling the identification of optimal or near-optimal settings. This methodology is particularly advantageous when resources are limited, as it provides a structured approach to experimental design. Unlike traditional methods, such as grid search, which exhaustively tests all combinations of hyperparameter levels which can be resource-intensive, RSM efficiently explores the response surface and captures potential curvature, leading to more effective optimization.

\section{Response Surface Methodology Based Hyperparameter Tuning}
   
Suppose we aim to minimize loss in a machine learning algorithm with $k$ hyperparameters. The complexity of the process prevents us from fitting a nonlinear model. However, we can model it as follows:
    \begin{equation}
        y_i=h(x_{1,i}, x_{2,i}, ..., x_{k, i})+\epsilon_i, i=1, 2, \hdots, n, 
    \end{equation}
where $x_{j,i}$ is the $i^{th}$ setting of the $j^{th}$ hyperparameter, and \(h(\cdot)\) is an unknown function that describes the relationship between the hyperparameters and their performance. This model is referred to as a response surface. The response surface is often approximated by a second-order regression model, since the main effects and second-order effects will essentially capture the nature of the response function and help identify the optimal response.

RSM is a DOE technique that identifies the curvature in the response surface with relatively few experiments. Modeling curvature effectively is of prime interest in hyperparameter optimization, since our objective is to identify the combination of hyperparameters that lead to the minimum of the loss function. RSM can improve the inefficiency of grid search by only using three levels per hyperparameter and two levels for its base design. 

To begin the optimization process, we must select the domain of each hyperparameter in our model. Proper domain selection involves setting reasonable ranges for each hyperparameter to ensure efficient exploration and to avoid wasted computational resources on suboptimal configurations. This process often starts with understanding the specific model and problem context, leveraging domain knowledge and empirical evidence. In this paper, we select the default hyperparameter setting given in \cite{Schelldorfer2019} as the midpoint of each domain and set the boundaries of the domain equidistant from the midpoint. This approach allows us to create a balanced search space that is centered around commonly used values while also permitting exploration of both higher and lower values, thereby increasing the likelihood of identifying an optimal configuration.

We will utilize the factorial arrangement DOE technique. A designed experiment is an exploration of groups and their responses within a specified framework. The experimenter obtains observations, measurements, and evaluations of the groups and conducts statistical inferences to provide reliable comparisons between them \cite{Ott2001}. By systematically varying the factors and analyzing the effects, we can determine the influence of each factor on the response variable, identify interaction effects, and optimize the process or product being studied. This approach allows for a thorough understanding of the relationships between variables and supports data-driven decision-making.

Here are some terminologies defined for DOE techniques:
\begin{itemize}
    \item Factor - Variables controlled by the experimenter for comparison;
    \item Factor level - The different values or settings that each factor in an experiment can take;
    \item Response - Measured or observed variables;
    \item Treatment - Conditions constructed by crossing the levels of each factor;
    \item Factorial design - All possible combinations of levels of multiple factors;
    \item Fractional factorial design - Carefully selected subsets of the full factorial design;
    \item Replication - Repetition of the same experimental conditions to obtain multiple observations.
\end{itemize}

For the $i^{th}$ level of the $j^{th}$ quantitative hyperparameter, we must use the following coding scheme to make the hyperparameters dimensionless:

\begin{equation}
        {x_{j,i}^\star=\frac{x_{j,i}-m_j}{s_j}=\frac{x_{j,i}-\frac{\underset{{1\leq i\leq n}}{\max}(x_{j,i}) + \underset{{1\leq i\leq n}}{\min}(x_{j,i})}{2}}{\frac{\underset{{1\leq i\leq n}}{\max}(x_{j,i})-\underset{{1\leq i\leq n}}{\min}(x_{j,i})}{2}}},\label{encoding}
\end{equation}
for every $j=1, 2, \ldots, k$ and $i=1, 2, \ldots, n$. This transformation encodes the low level of factor $j$ to a coded value of $-1$, the high level of factor $j$ to a coded value of 1, and the mid level of factor $j$ to a coded value of 0. When every factor is set at its mid level, $m_j$, in a treatment, it represents the center of the experiment, suspected to be the optimal region.

Since the location of the optimal region of hyperparameters is typically unknown, a sequential search strategy is employed. This strategy involves conducting three separate trials to identify the optimal or near-optimal set of hyperparameters.

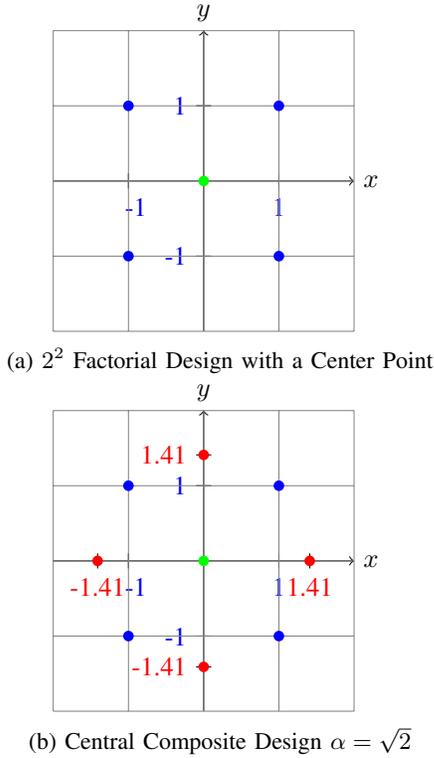
\begin{figure}[htbp]
    \centering
    \begin{subfigure}[b]{0.45\textwidth}
        \centering
        \begin{tikzpicture}[scale=1]
            \draw[->] (-2,0) -- (2,0) node[right] {$x$};
            \draw[->] (0,-2) -- (0,2) node[above] {$y$};

            \draw (-1, 0.1) -- (-1, -0.1) node[below] {\hspace{0.2cm}\textcolor{blue}{-1}};
            \draw (1, 0.1) -- (1, -0.1) node[below] {\textcolor{blue}{1}};

            \draw (0.1, -1) -- (-0.1, -1) node[left] {\textcolor{blue}{-1}};
            \draw (0.1, 1) -- (-0.1, 1) node[left] {\textcolor{blue}{1}};

            \draw[help lines] (-2,-2) grid (2,2);

            \fill[fill=blue] (1, 1) circle (2pt);
            \fill[fill=blue] (1, -1) circle (2pt);
            \fill[fill=blue] (-1, -1) circle (2pt);
            \fill[fill=blue] (-1, 1) circle (2pt);
            \fill[fill=green] (0, 0) circle (2pt);
        \end{tikzpicture}
        \caption{$2^2$ Factorial Design with a Center Point}
    \end{subfigure}
    \hfill
    \begin{subfigure}[b]{0.45\textwidth}
        \centering
        \begin{tikzpicture}[scale=1]
            \draw[->] (-2,0) -- (2,0) node[right] {$x$};
            \draw[->] (0,-2) -- (0,2) node[above] {$y$};

            \draw (-1, 0.1) -- (-1, -0.1) node[below] {\hspace{0.2cm}\textcolor{blue}{-1}};
            \draw (1, 0.1) -- (1, -0.1) node[below] {\textcolor{blue}{1}};
            \draw (1.41, 0.1) -- (1.41, -0.1) node[below] {\textcolor{red}{1.41}};
            \draw (-1.41, 0.1) -- (-1.41, -0.1) node[below] {\textcolor{red}{-1.41}};

            \draw (0.1, -1) -- (-0.1, -1) node[left] {\textcolor{blue}{-1}};
            \draw (0.1, 1) -- (-0.1, 1) node[left] {\textcolor{blue}{1}};
            \draw (0.1, 1.41) -- (-0.1, 1.41) node[left] {\textcolor{red}{1.41}};
            \draw (0.1, -1.41) -- (-0.1, -1.41) node[left] {\textcolor{red}{-1.41}};

            \draw[help lines] (-2,-2) grid (2,2);

            \fill[fill=blue] (1, 1) circle (2pt);
            \fill[fill=blue] (1, -1) circle (2pt);
            \fill[fill=blue] (-1, -1) circle (2pt);
            \fill[fill=blue] (-1, 1) circle (2pt);
            \fill[fill=green] (0, 0) circle (2pt);
            \fill[fill=red] (1.41, 0) circle (2pt);
            \fill[fill=red] (-1.41, 0) circle (2pt);
            \fill[fill=red] (0, 1.41) circle (2pt);
            \fill[fill=red] (0, -1.41) circle (2pt);
        \end{tikzpicture}
        \caption{Central Composite Design $\alpha=\sqrt{2}$}
    \end{subfigure}

    \caption{Two Factor Response Surface Design}
    \label{fig:twofactorrRSD}
\end{figure}

First, a screening design is performed across specified hyperparameters and their domains. This initial design uses 3 levels per hyperparameter. However, the first $2^k$ design points are set at 2 levels per hyperparameter, specifically $\left\{\underset{{1\leq i\leq n}}{\max}(x_{j,i}), \underset{{1\leq i\leq n}}{\min}(x_{j,i})\right\}$, followed by a center point where every hyperparameter is at its mid level. As an example, consider Figure~\ref{fig:twofactorrRSD}~(a), which employs a $2^2$ factorial design where $k=2$ with a center point. The response surface over the coded values in this initial design is modeled using a first-order regression model given by:
\begin{equation}
\begin{aligned}
E[Y] &= \beta_0 + \beta_1X_1^\star + \ldots + \beta_kX_k^\star.\label{FirstOrdModel}
\end{aligned}
\end{equation}

Based on the results, less significant hyperparameters may be dropped and set at their midlevels. The gradient of the first-order model is then used to estimate the direction towards which minimum loss is likely located. This direction, known colloquially as the path of steepest descent (or ascent), guides additional experiments along this path with a step length \(t\) to identify the new optimal region of interest.

\begin{equation}
    X_h = \frac{t}{s} \mathbf{b^{\star}} \label{SteepestDescent}
\end{equation}
where

\begin{equation*}
    \mathbf{b^{\star}} = 
    \begin{bmatrix}
        b_1 & b_2 & \ldots & b_k
    \end{bmatrix}^T
    \quad \text{and} \quad 
    s = \sqrt{b_1^2 + \ldots + b_k^2}.
\end{equation*}

Since we are following the steepest descent path, \(t\) is negative; otherwise, it is positive. The new design points generated by \eqref{SteepestDescent} are in their coded units. Thus, appropriate decoding must be done using the same \(m_j\) and \(s_j\) by

\begin{equation}
    x_{j,i} = \left\{ 
    \begin{array}{rcl}
        x_{j,i}^\star s_j + m_j & \text{for} & X_j \in \mathbb{R} \\ 
        \nint{x_{j,i}^\star s_j + m_j} & \text{for} & X_j \in \mathbb{N}.
    \end{array} 
    \right. \label{Decoding}
\end{equation}

After we have identified the new optimal center of the experiment, $C=(c_1, c_2, \ldots, c_p)$ where $p\leq k$, we define a new domain $X_j\in\{x_{j,i}: |x_{j,i}-c_j|\leq d\}$ for some $d\in \mathbb{R}$. This design uses three levels per hyperparameter. However, the first $2^p$ design points are set at two levels for every hyperparameter, specifically $\left\{\underset{{1\leq i\leq n}}{\max}(x_{j,i}), \underset{{1\leq i\leq n}}{\min}(x_{j,i})\right\}$, followed by a center point where every hyperparameter is at its mid level, and $2p$ star points are added. A star point is a point with all the factors except one set at their mid-levels. Every hyperparameter has two star points, ${S_{1,j}, S_{2,j}}$, that are the opposites of each other, i.e., $S_{2,j}=-S_{1,j}$ for every $j=1, \hdots, p$. The coordinates of the star points in Figure~\ref{fig:twofactorrRSD} (b) are $(\pm\sqrt{2}, 0)$ and $(0, \pm\sqrt{2})$. 

This design is called a Central Composite Design (CCD). CCD is a full or fractional factorial design with two levels on each factor that have an additional small number of treatments to permit estimation of the second-order response surface model given by

\begin{equation}
\begin{aligned}
E[Y] &= \beta_0 + \beta_1X_1 + \ldots + \beta_pX_p \\
     &\quad + \beta_{11}X_1^2 + \ldots + \beta_{pp}X_p^2 \\
     &\quad + \beta_{12}X_1X_2 + \ldots + \beta_{p-1,p}X_{p-1}X_p\label{SecOrdMod}
\end{aligned}
\end{equation} where

\begin{itemize}
    \item $\beta_1, ..., \beta_p$ – linear main effects
    \item $\beta_{11}, ..., \beta_{pp}$ – quadratic effects
    \item $\beta_{12}, \beta_{13}, ..., \beta_{p-1,p}$ – interaction effects.
\end{itemize} 

CCDs have been developed for estimating response surfaces based on the second order model \eqref{SecOrdMod} to overcome the limitations of designs with more than two factor levels in which the number of experiments grows exponentially. In CCD, we will run $2^{p-f}n_c+2pn_s+n_0$ number of trials where
\begin{itemize}
    \item $p$ - number of hyperparameters
    \item $f$ – level of fractionation for fractional factorial designs
    \item $n_c$ – number of replication at each corner point
    \item $n_s$ – number of replications at each star point
    \item $n_0$ - number of replications at the center point.
\end{itemize}

However, when conducting CCD, we must consider the precision of our estimate $E[Y_h]$  at different values of the predictor variables, $X_h$. This necessitates the rotatability criterion. An experimental design is rotatable if "the variance of the fitted value at $X_h, \sigma^2{\hat{Y_h}}$, is the same for any point $X_h$ that is a given distance from the center point, regardless of the direction \cite{Kutner2005} and It can be shown that a CCD is rotatable if the coordinates of the nonzero entry of every star point are
\begin{equation}
    \alpha=\pm\left[\frac{2^{p-f}n_c}{n_s}\right]^{1/4}.
    \label{Star Point Distance}
\end{equation}

\section{Related Works}

Over the last few decades, numerous hyperparameter tuning techniques have been proposed. A simple approach is to choose the default hyperparameters recommended by the algorithm developers or found in the available literature. However, since each data set is different, these hyperparameter choices may not perform well on various data sets or even on different devices. Alternatively, grid search (GS) can be employed to sift through each combination of hyperparameters to find the best one for the task from a given range of hyperparameters and step sizes. However, this method is time-consuming and often inefficient. We may miss the global optimum (or near global optimum) if we simply vary the variables one-by-one, as the maximal value of one variable is usually dependent on the others.

A more sophisticated approach considers hyperparameter tuning as an optimization problem, aiming to find the function that maximizes model performance or minimizes errors \cite{Pannakkong2022}. Examples of such techniques include Bayesian optimization \cite{Frazier2018}, gradient-based optimization \cite{Bakhteev2020}, genetic algorithms \cite{Syarif2016}, and surrogate models \cite{abraham2023ncqs,abraham2023homopt}. However, these techniques can be challenging for inexperienced modelers and often require a large number of runs or substantial memory resources. This creates a demand for an easy-to-use, yet efficient tuning strategy that reduces the number of runs required to find the optimal hyperparameters.

Reference \cite{Pannakkong2022} showed that when Response Surface Methodology (RSM) is applied to artificial neural networks (ANN), support vector machines (SVM) and deep belief networks (DBN), it produces greater precision than the grid search approach. In addition, savings in the number of GS runs are 97. 79\%, 97. 81\%, and 80. 69\% for ANN, SVM, and DBN, respectively. Reference \cite{Moreno2018} found that RSM, when applied to the random forest classification algorithm, was overall better than GS and reduced the number of runs required from 16,384 to 157.

\section{Model Description and Data Set}

\subsection{Combined Actuarial Linear Model}

\begin{figure}[htbp]
\tikzset{%
  every neuron/.style={
    circle,
    draw,
    minimum size=0.4cm
  },
  neuron missing/.style={
    draw=none, 
    scale=4,
    text height=0.333cm,
    execute at begin node=\color{black}$\vdots$
  },
  point/.style={
    fill,
    circle,
    inner sep=0.1pt
  }
}

\begin{tikzpicture}[x=0.8cm, y=0.6cm, >=stealth]

\foreach \m/\l [count=\y] in {1,2,3,4,5,6,7,8,9,10,11}
  \node [every neuron/.try, neuron \m/.try] (input-\m) at (0, 5.5-\y) {};

\node [every neuron] (input-12) at (0, 6.5-15) {};

\foreach \m [count=\y] in {1,2,3,4,5,6,7,8,9,10,11,12,13,14,15,16,17,18,19,20}
  \node [every neuron/.try, neuron \m/.try ] (hidden1-\m) at (2, 10-\y) {};

\node [point] (hidden1-21) at (2, 11-22) {};

\foreach \m [count=\y] in {1,2,3,4,5,6,7,8,9,10,11,12,13,14,15}
  \node [every neuron/.try, neuron \m/.try ] (hidden2-\m) at (4, 7.5-\y) {};

\node [point] (hidden2-16) at (4, 8.5-17) {};

\foreach \m [count=\y] in {1,2,3,4,5,6,7,8,9,10}
  \node [every neuron/.try, neuron \m/.try ] (hidden3-\m) at (6, 5-\y) {};

\node [point] (hidden3-11) at (6, 6-12) {};

\foreach \m [count=\y] in {1}
  \node [every neuron/.try, neuron \m/.try ] (output-\m) at (8, 2.5-\y) {};

\node [circle] (output-2) at (8, 3.5-2) {};

\foreach \i in {1,2,...,11}
  \foreach \j in {1,2,...,20}
    \draw [->] (input-\i) -- (hidden1-\j);

\foreach \i in {1,2,...,20}
  \foreach \j in {1,2,...,15}
    \draw [->] (hidden1-\i) -- (hidden2-\j);

\foreach \i in {1,2,...,15}
  \foreach \j in {1,2,...,10}
    \draw [->] (hidden2-\i) -- (hidden3-\j);

\foreach \i in {1,2,...,10}
  \draw [->] (hidden3-\i) -- (output-1);

\draw [-, red] (input-12) -- (hidden1-21);
\draw [-, red] (hidden1-21) -- (hidden2-16);
\draw [-, red] (hidden2-16) -- (hidden3-11);
\draw [->, red] (hidden3-11) -- (output-2);

\node at (-0.5, 6.5-14.25) {GLM};

\node [align=center, above] at (0, 6) {Input\\layer};
\node [align=center, above] at (2, 9.5) {Hidden\\layer 1};
\node [align=center, above] at (4, 8) {Hidden\\layer 2};
\node [align=center, above] at (6, 5.5) {Hidden\\layer 3};
\node [align=center, above] at (8, 3) {Output\\layer};

\end{tikzpicture}
\caption{Neural Network architecture with $Lr=3$ hidden layers (20, 15, 10 neurons respectively) and GLM with skip connection}
\label{fig:NNA}
\end{figure}
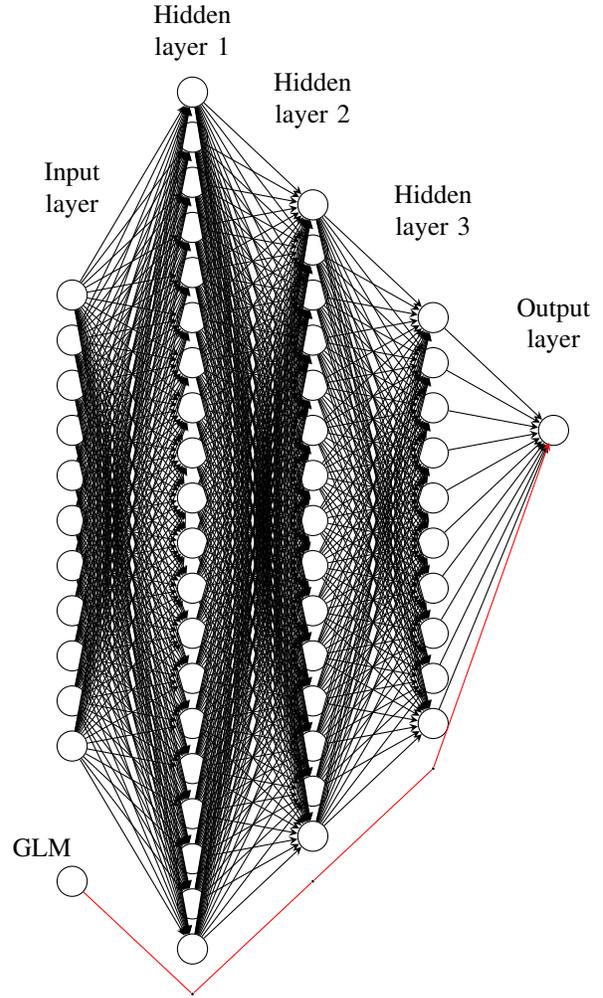

In this section, we briefly introduce the CANN model. CANN combines traditional actuarial models, such as Generalized Linear Models (GLM), with neural networks to enhance predictive accuracy. We optimize the CANN architecture presented by \cite{Schelldorfer2019}, which embeds a classical generalized linear model into a feedforward neural network (FFNN) via a skip connection linking the input layer to the output neuron, as illustrated in Figure~\ref{fig:NNA} in red. This skip connection is implemented through an addition in the output layer. Although its original CANN implementation was in R, we have translated it to Python using Keras and TensorFlow on a 2019 MacBook Pro.

\subsubsection{CANN Model Assumptions}

Choose a feature space $\mathcal{X}\subset \mathbb{R}^{q_0}$  and define regression function $\lambda:\mathcal{X}\mapsto\mathbb{R}_+$ by
\begin{equation}
    \mathbf{x}\mapsto log\lambda(\mathbf{x})=\langle \beta,\mathbf{x}\rangle+\langle \mathbf{w}^{(Lr+1)},(\mathbf{z}^{(Lr)}\circ ... \circ\mathbf{z}^{(1)})(\mathbf{x})\rangle,
\end{equation}
where the first term on the right-hand side is the GLM regression function with parameter vector $\beta=(\beta_0, \hdots, \beta_{q_0})^\intercal\in\mathbb{R}^{q_0+1}$ given by 
\small{
\begin{equation}
    \begin{split}
        \langle \beta,\mathbf{x}\rangle &= \beta_0 + \sum_{l=1}^{q_0}\beta_lx_l + \beta_{q_0+1}DrivAge \\
        &\quad + \beta_{q_0+2}\log(DrivAge) + \sum_{z=2}^4\beta_{q_0+z+1}(DrivAge)^z.
    \end{split}
    \label{GLM_Def}
\end{equation}} 

The second term is the Neural Network regression function with number of layers equal to $Lr \in\mathbb{N}$ and with $q_0$ dimensional input layer. Assume that the number of claims $N_i$ for a given policy $i$ is independent and Poisson distributed, $N_i\sim$ Poi$(\lambda(\mathbf{x}_i)\nu_i)$ for all $i\geq1$.

\subsection{French Motor Third Party Liability Insurance Data}\label{AA}
\begin{figure}[h]
    \centering
    \includegraphics[width=1\linewidth]{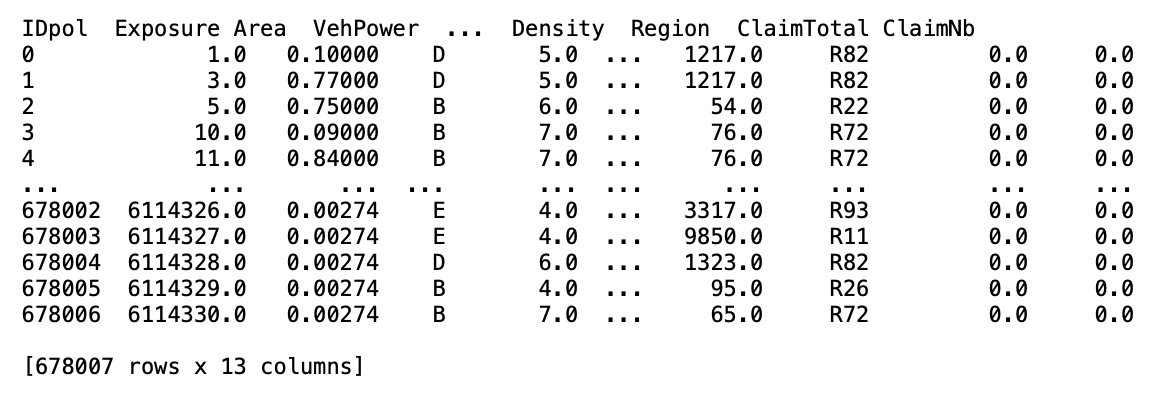}
    \caption{freMTPL2freq Summary}
    \label{fig:freMTPL2freq_summary}
\end{figure}

Our computational experiments are conducted on the French motor third-party liability insurance portfolio data set, freMTPL2freq, which is included in the R package CASdatasets \cite{freMTPL2freq}. This data set contains the French MTPL insurance portfolio with the corresponding claim counts observed within one accounting year. Figure~\ref{fig:freMTPL2freq_summary} provides a short summary of the data.

Reference \cite{Schelldorfer2019} details the precautions necessary for handling attributes in this data set, which we will not discuss further.

This data set comprises 678,007 insurance policies. As described in \cite{Schelldorfer2019}, we assume that the number of claims \(N_i\) for a given policy \(i\) is independent and Poisson distributed with

\begin{equation}
    N_i \sim \mathrm{Poi}(\lambda(\mathbf{x}_i)\nu_i),
\end{equation}

where \(\nu_i > 0\) represents the given volumes (Exposure in yearly units) and \(\mathbf{x}_i \mapsto \lambda(\mathbf{x}_i)\) is the regression function, with \(\mathbf{x}_i\) being the feature vector of policy \(i\).

freMTPL2freq has 12 attributes that are:
\begin{itemize}
    \item IDpol - policy number;
    \item ClaimNb - number of claims on a given policy;
    \item Exposure - total exposure in yearly units;
    \item Area - area code;
    \item VehPower - power of the car;
    \item VehAge - age of the car in years;
    \item DrivAge - the most common age of the driver in years;
    \item BonusMalus - bonus-malus level between 50 and 230;
    \item VehBrand - car brand;
    \item VehGas - car fuel type (diesel or regular);
    \item  Density - density of inhabitants per $km^2$
in the living place of the driver;
    \item Region: regions in France (prior to 2016).
\end{itemize}

Our goal is to optimize the hyperparameters of the CANN model used to estimate the regression function \(\lambda(\cdot)\).

We will use the feature extraction method performed by \cite{Schelldorfer2019}.

\subsection{Data Processing for GLM}
We will perform the following data preprocessing for the GLM part of the CANN:

\begin{itemize}
    \item Area: \{A, ..., F\}$\mapsto$\{1, ..., 6\};
    \item VehPower: 2 categorical classes by $(-\infty, 9], (9, \infty)$;
    \item VehAge: 3 categorical classes $[0, 1), [1, 10], (10, \infty)$;
    \item DriveAge: 7 categorial classes $[18, 21), [21, 26), [26, 31)$, $[31, 41), [41, 51), [51, 71), [71, \infty)$ and a functional form 
    \begin{center}
        DrivAge $\mapsto\beta_{q_0+1}$DrivAge$ + \beta_{q_0+2}\log(\mathrm{DrivAge}) + \sum_{z=2}^4\beta_{q_0+z+1}(\mathrm{DrivAge})^z$;
    \end{center}
    \item BonusMalus: Continuous log-linear feature component capped at 150;
    \item VehBrand: 11 categorical classes;
    \item VehGas: Binary feature component;
    \item Density: Continuous feature component log-density;
    \item Region: 22 categorical feature component.
\end{itemize}

\subsection{Data Processing for FFNN}
We will perform the following data preprocessing for the FFNN part of the CANN:
\begin{itemize}
    \item Continuous features $x_l\mapsto x_l^\star=\frac{2(x_l-min_l)}{Max_l-min_l}-1\in[-1, 1]$;
    \item Binary features set to $\{-1/2, 1/2\}$;
    \item Categorical features $\mapsto$ dummy coding $\mapsto$ 2D embedding.
\end{itemize}

\section{Computational Experiments and Setup}
\subsection{CANN Implementation}
In the CANN implementation, we randomly allocate $80\%$ of our 678007 data points to the training set $\mathcal{D}$ and the remaining $20\%$ of the data to the testing set $\mathcal{T}$.

    We then fit the CANN model on the learning data set $\mathcal{D}$ by minimizing the corresponding \textit{in-sample Poisson deviance loss function} 
    \begin{center}
        \centering
        $\mathbf{\beta} \mapsto \mathcal{L}(\mathcal{D}, \lambda)=\frac{1}{n}\sum_{i=1}^n 2N_i [\frac{\lambda (\mathbf{x_i})\nu_i}{N_i}-1-log(\frac{\lambda (\mathbf{x_i})\nu_i}{N_i})]$
    \end{center}
    for $\mathbf{\beta}$-dependent parametric regression function $\lambda (\cdot)=\lambda_{\beta}(\cdot)$, and the summation runs over all policies $1\leq i \leq n_D=542405$ in the training data set $\mathcal{D}$. Denote the resulting maximum likelihood estimate (MLE) by $\hat{\beta}$. This provides the estimated regression function $\hat{\lambda}(\cdot)=\lambda_{\hat{\beta}}(\cdot)$.

    The quality of this model is assessed on the \textit{out-of-sample Poisson deviance loss} (generalization loss) on the test data set $\mathcal{T}$ given by 
    \begin{center}
        \centering
        $\mathcal{L}(\mathcal{D}, \hat{\lambda})=\frac{1}{n_{\mathcal{T}}}\sum_{t=1}^{n_{\mathcal{T}}} 2N_t [\frac{\hat{\lambda} (\mathbf{x_t})\nu_t}{N_t}-1-log(\frac{\hat{\lambda} (\mathbf{x_t})\nu_t}{N_t})]$,
    \end{center}
    where the summation runs through all policies $1 \leq t \leq n_{\mathcal{T}}=135601$ in the test data set $\mathcal{T}$.

In this implementation of CANN, the following hyperparameters are chosen to be optimized.

    \begin{itemize}
        \item 7 levels of optimizer: \{Adagrad, Adadelta, SGD, RMSprop, Adam, Adamax, Nadam\}$\mapsto\{0, 1, \hdots, 6\}$, Op;
        \item Number of neurons in the hidden layers, $(N1, N2, N3)$;
        \item Number of epochs, Ep;
        \item Batch size, Bh;
        \item Number of hidden layers, Lr.
    \end{itemize}

\subsection{Initial Experiments}
To start our experiments, we must define the range of values for each hyperparameter. 
By setting the default hyperparameters used by \cite{Schelldorfer2019} as mid level, we obtain the following ranges:
    \begin{itemize}
        \item Optimizer, Op $\in [0, 6]$
        \item Number of neurons in the hidden layers: $N1 \in [10, 30], N2 \in [5, 25], N3 \in [5, 15]$
        \item Epoch, Ep $\in [100, 900]$
        \item Batch size, Bh $\in [5000, 15000]$;
        \item Layer, Lr $\in [2, 4]$
    \end{itemize}
Our initial design points consist of $2^7$ factorial arrangement treatments with minimum and maximum values as their two levels for every hyperparameter. An additional center point is added where each hyperparameter is at its mid-level, $\frac{\underset{{1\leq i\leq n}}{\max}(x_{j,i}) + \underset{{1\leq i\leq n}}{\min}(x_{j,i})}{2}$ for each $j = 1, \ldots, 7$, with 4 replications. We will use out-of-sample poisson deviance loss as the response variable.

The experimental result is obtained after 132 runs, and a first-order regression model is fit over the coded values.
The regression parameter estimates and the intercept are shown in Table \ref{tab:FirstOrderParameters}.

\begin{table}[htbp]
    \centering
    \caption{First Order Regression Model}
    \begin{tabular}{|c|c|c|c|c|c|c|}
        \hline
        Variable & Parameter & STD Error & t Value & P-value \\
        \hline
        Intercept & 46.0791 & 2.83375 & 16.26 & 0.0001 \\
        \hline
        Optimizer & -21.8703 & 2.87768 & -7.60 & 0.0001 \\
        \hline
        N1 & -0.5261 & 2.8777 & -0.18 & 0.8552 \\
        \hline
        N2 & -11.0132 & 2.8777 & -3.83 & 0.0002 \\
        \hline
        N3 & -0.9574 & 2.8777 & -0.33 & 0.7399 \\
        \hline
        Epoch & -16.005 & 2.8777 & -5.56 & 0.0001 \\
        \hline
        Batch & 9.1987 & 2.8777 & 3.20 & 0.0018 \\
        \hline
        Layer & -7.3396 & 2.8777 & -2.55 & 0.012 \\
        \hline
    \end{tabular}
    \label{tab:FirstOrderParameters}
\end{table}

\subsection{Path of Steepest Descent}

The gradient of this first-order regression model is calculated and \eqref{SteepestDescent} is used to generate 20 additional design points. Of these, 10 points utilize all the hyperparameters, while the other 10 points hold the number of neurons in the $1^{\text{st}}$ and $3^{\text{rd}}$ hidden layers constant at their midlevels, $N1=20$ and $N3=15$. Moreover, with every additional hidden layer, we will set the number of neurons of that layer default to 15. This approach aims to assess whether there is a significant difference between tuning all hyperparameters versus omitting less significant ones. By comparing the performance across these two sets of design points, we can evaluate the impact of excluding certain hyperparameters on the overall model optimization. This can help determine if reducing the complexity of the hyperparameter tuning process by focusing on the most impactful parameters leads to comparable or improved model performance, thereby potentially saving computational resources and simplifying the tuning process.

Note that during the descent, we encountered a problem with the optimizer values going out of bounds after decoding. This issue was remedied by taking $x_{j,1}'\equiv x_{j,1}\mod 7$. This approach ensures that the values remain within the acceptable range, effectively preventing any instability or errors that might arise from out-of-bound values during the optimization process. Additionally, since the optimizer is essentially a nominal attribute, rotating the optimizers used will better explore the effects of different optimizers.

Table \ref{tab:OptimalPoints} presents the new optimal hyperparameter settings discovered through the steepest descent path in both complete and reduced hyperparameter tuning, together with their corresponding out-of-sample Poisson deviance losses. Interestingly, the reduced hyperparameter tuning captured the same optimal hyperparameter levels as the complete hyperparameter tuning.

\begin{table}[htbp]
    \centering
    \caption{New Optimal Points}
    \begin{tabular}{|c|c|c|c|c|c|c|c|c|}
        \hline
         HP & Op & N1 & N2 & N3 & Ep & Bh & Lr & Loss\\
        \hline
        Complete & 5 & 20 & 18 & 10 & 703 & 8542 & 3 & 0.24598\\
        \hline
        Reduced & 5 & 20 & 18 & 10 & 703 & 8542 & 3 & 0.24598\\
        \hline
    \end{tabular}
    \label{tab:OptimalPoints}
\end{table}

\subsection{Second Order Experimental Design Setup}

Next, we will define new boundaries for each hyperparameter to be tuned and set the new optimal point found by the steepest descent as the CCD midpoint.

\begin{table}[htbp]
    \centering
    \caption{Second Order Design Hyperparameter Domains}
    \begin{tabular}{|c|c|c|}
        \hline
         HP & Complete & Reduced\\
        \hline
        Op & [4, 6] & [4, 6]\\
        \hline
        N1 & [15, 25] & 20\\
        \hline
        N2 & [13, 23] & [13, 23]\\
        \hline
        N3 & [8, 12] & 10\\
        \hline
        Ep & [503, 903] & [503, 903]\\
        \hline
        Bh & [6542, 10542] & [6542, 10542]\\
        \hline
        Lr & [2, 4] & [2, 4]\\
        \hline
    \end{tabular}
    \label{tab:SecondOrderDesignHyperparanterDomains}
\end{table}

The design points in these two CCDs include $2^7$ and $2^5$ factorial base designs, encompassing two levels that are the maximum and minimum values of each hyperparameter. Additionally, each design incorporates center points with 4 replications. There are also 14 star points with non-zero entries $x_{j,s_j}^{\star}=\pm 2^{7/4}$ and 10 star points with non-zero entries $x_{j,s_j}^{\star}=\pm 2^{5/4}$ , determined using \eqref{Star Point Distance} for every $j=1, \hdots, p$. After appropriate decoding is performed using \eqref{Decoding} and the new $m_j'$ and $s_j'$, the associated levels are located in the actual units of the hyperparameters. This experimental setup results in 146 runs for the complete hyperparameter tuning and 46 runs for the reduced hyperparameter tuning designs for the second-order model CCD.

Note that while we were decoding star points with non-zero optimizer entries, we encountered out-of-bounds values $\nint{x_{1,s_1}}=\nint{2^{7/4}*1+5}\approx\nint{8.3636}=8$ and $\nint{x_{1,s_2}}=\nint{2^{5/4}*1+5}\approx\nint{7.3784}=7$ for both the complete and the reduced hyperparameter cases respectively. To handle such situations, we propose two different methods and a criterion for determining which design is more suitable. We could replace the out-of-bound values with the closest in-range hyperparameter level, or we could rotate the hyperparameters by taking modulo $q$ where $q$ is the number of optimizers available. This approach ensures that the hyperparameter values remain within a specified range, addressing the issue of out-of-bound values encountered during the optimization process.

To determine which method is better, we introduce the \textit{D} criterion. As discussed in \cite{Kutner2005}, when precise estimation of model parameters is needed, the \textit{D} criterion provides a measure of the precision of the experiment. For regression models with multiple parameters, the confidence region is an ellipsoid. Thus, we need to find the area or the volume of this region, since a smaller confidence region implies higher precision. The \textit{D} criterion achieves this by minimizing 
\begin{equation}
    D=|(X^\intercal X)^{-1}|
    \label{D_criterion}
\end{equation}
where $X$ is the design matrix of predictors in coded units.

However, we decided to examine both cases of handling star points for both the complete and the reduced hyperparameter designs. This approach allows us to compare the effectiveness and performance impacts of different methods for handling irregular experimental regions while acknowledging the fact that, in Table \ref{tab:CriterionDeterminantResults}, the rotated optimizer values attained the minimum determinant in both the complete and reduced hyperparameter cases.

\begin{table}
    \centering
    \caption{\textit{D} Criterion Determinant Results}
    \begin{tabular}{|c|c|c|}
    \hline
          & Closest & Modulo\\
        \hline
        Complete & 6.10e-16 & 5.51e-16\\
        \hline
        Reduced & 7.35e-9 & 4.53e-9\\
        \hline
    \end{tabular}
    \label{tab:CriterionDeterminantResults}
\end{table}

\subsection{Second Order Experimental Results and Analysis}

In this final stage of the tuning process, we fit a second-order regression model \eqref{SecOrdMod} on the results we have obtained. The optimal point where a maximum, a minimum, or a saddle point occurs in \eqref{SecOrdMod}, denoted $X_o$, is calculated by:
\begin{equation}
    X_o=-\frac{1}{2}\textbf{B}^{-1}\textbf{b}^\ast \label{OptimalPointCalc}
\end{equation}
where
\begin{equation}
    \textbf{B} =
    \begin{bmatrix}
        b_{11} & \frac{b_{12}}{2} & \ldots & \frac{b_{1p}}{2} \\
        \frac{b_{12}}{2} & b_{22} & \ldots & \frac{b_{2p}}{2} \\
        \vdots & \vdots &   & \vdots \\
        \frac{b_{1p}}{2} & \frac{b_{2p}}{2} & \ldots & b_{pp}
    \end{bmatrix} \quad\quad
    \textbf{b}^\ast=
    \begin{bmatrix}
        b_1 \\
        b_2 \\
        \vdots \\
        b_p
    \end{bmatrix}.\label{OptimalPointCalcTerms}
\end{equation} 
Here, \textbf{B} is the regression parameter matrix of the quadratic and interaction terms of \eqref{SecOrdMod} and \textbf{b} is the vector of linear terms of \eqref{SecOrdMod}. The eigenvalues of \textbf{B} can be used to determine the shape of the response surface.
\begin{itemize}
    \item If the eigenvalues are all positive, then it is a minimum;
    \item If the eigenvalues are all negative, then it is a maximum;
    \item If the eigenvalues have mixed signs, then it is a saddle point.
\end{itemize}

We conduct this analysis using the actual units of the hyperparameters to obtain the predicted optimal hyperparameter sets and the estimated out-of-sample Poisson deviance losses, as shown in Table \ref{tab:2ndOrderPred}.

\begin{table}[htbp] 
    \centering
    \caption{Optimal Design Points and Predictions}
    \begin{tabular}{|c|c|c|c|c|c|c|c|c|}
        \hline
         HP & Op & N1 & N2 & N3 & Ep & Bh & Lr & $\underset{\mathrm{Loss}}{\mathrm{Predicted}}$\\
        \hline
        $\underset{\mathrm{Max}}{\mathrm{Complete}}$ & 5 & 20 & 19 & 10 & 726 & 8318 & 3 & 0.247589\\
        \hline
        $\underset{\mathrm{Mod}}{\mathrm{Complete}}$ & 5 & 20 & 17 & 10 & 723 & 8988 & 3 & 0.243246\\
        \hline
        $\underset{\mathrm{Max}}{\mathrm{Reduced}}$ & 4 & 20 & 9 & 10 & 650 & 6174 & 3 & 0.245938\\
        \hline
        $\underset{\mathrm{Mod}}{\mathrm{Reduced}}$ & 5 & 20 & 17 & 15 & 661 & 8783 & 3 & 0.241764\\
        \hline
    \end{tabular}
    \label{tab:2ndOrderPred}
\end{table}

Based on the results, we found that when out-of-bounds optimizer values were rotated by modulo \(n\), the response surfaces were minimized. However, when out-of-bounds optimizer values were set in the nearest in-range setting, the response surfaces exhibited saddle points in both the complete and reduced hyperparameter cases. 

We conducted confirmatory experiments on the suggested minimum and saddle points and obtained the following results, as shown in Table \ref{tab:2ndOrderObs}, together with the minimum response encountered during each tuning experiment. We observed that when we optimized every hyperparameter and rotated the optimizer values, we achieved a minimum loss that was better than any loss observed during experiments. However, in other instances, our predicted points were suboptimal. In such cases, we should repeat the process until we reach the best result or choose the best hyperparameters encountered during the experiments.

\begin{table}[htbp] 
    \centering
    \caption{Optimal Points Confirmation Runs}
    \begin{tabular}{|c|c|c|c|}
        \hline
           & $\underset{\mathrm{Loss}}{\mathrm{Predicted}}$ & $\underset{\mathrm{Loss}}{\mathrm{Observed}}$ & $\underset{\mathrm{Min}}{\mathrm{Historic}}$\\
        \hline
        $\underset{\mathrm{Max}}{\mathrm{Complete}}$ & 0.247589 & 0.245960 & 0.245886\\
        \hline
        $\underset{\mathrm{Mod}}{\mathrm{Complete}}$ & 0.243246 & 0.245823 & 0.245886\\
        \hline
        $\underset{\mathrm{Max}}{\mathrm{Reduced}}$ & 0.245938 & 0.247471 & 0.245976\\
        \hline
        $\underset{\mathrm{Mod}}{\mathrm{Reduced}}$ & 0.241764 & 0.246105 & 0.24.5976\\
        \hline
    \end{tabular}
    \label{tab:2ndOrderObs}
\end{table}

\section{Future Directions}

While this study has demonstrated the effectiveness of Response Surface Methodology (RSM) in hyperparameter optimization for neural networks in actuarial applications, there are several promising avenues for future research and development.

Firstly, the integration of surrogate models could further enhance the efficiency of hyperparameter optimization. Surrogate models, such as nonlinear convex quadrature surrogate hyperparameter optimization (NCQS) \cite{abraham2023ncqs}, can approximate the objective function and provide insights with significantly fewer evaluations. These models can be particularly useful when dealing with expensive objective functions, as they allow for a more informed exploration of the hyperparameter space.

Secondly, homotopy-based approaches \cite{abraham2023homopt} offer another exciting direction. Homotopy methods, which gradually transform a simple problem into a more complex one, can be employed to guide the optimization process. Using homotopy techniques, it is possible to explore the hyperparameter space more effectively and avoid local minima, thus improving the robustness of the optimization process.

In addition, combining these advanced techniques with RSM could lead to a hybrid approach that takes advantage of the strengths of each method. For instance, surrogate models could be used to identify promising regions of the hyperparameter space, which can then be refined using RSM. Similarly, homotopy methods can be integrated with RSM to provide a smoother and more continuous search path, enhancing the overall optimization process.

Lastly, extending these methodologies to other machine learning algorithms beyond neural networks could provide valuable insights and benefits across various domains. Techniques like support vector machines, gradient boosting machines, and ensemble methods could also benefit from the structured and efficient hyperparameter optimization strategies discussed in this study.

In summary, future research should focus on exploring these advanced techniques and their combinations to further improve hyperparameter optimization. By doing so, we can continue to enhance the performance and efficiency of machine learning models in actuarial science and beyond.

\section{Conclusion}

Hyperparameter optimization is a crucial step in implementing any machine learning algorithm. Recent advances in the application of neural networks in actuarial science have demonstrated significant benefits by integrating these advanced computational techniques into the traditional actuarial toolbox. Studies have shown that neural network predictions have surpassed traditional actuarial methods in areas such as pricing \cite{Schelldorfer2019}, claims reserving \cite{Gabrielli2018}, and mortality forecasting \cite{Hainaut2018}.

To fully leverage the potential of neural networks, various hyperparameter optimization methods have been proposed. Given that these methods often require a large number of iterations, RSM provides a structured approach to reduce the number of necessary iterations while minimizing the loss function.

We have demonstrated that RSM accurately predicts optimal hyperparameter combinations with significantly fewer runs required compared to GS on four different cases of optimization a modeler might encounter. The total runs needed for RSM is calculated as follows:
\begin{equation}
    T_{RSM}=2^kn_c+n_{0_1}+n_t+2^{p-f}n_c'+2pn_s+n_{0_2}
\end{equation}
\begin{itemize}
    \item $n_{0_i}$ - number of replications of the center point for each phase of the experiment, $i=1,2$;
    \item $n_t$ - number of new points generated by the path of steepest descent.
    \item $n_c, n_c'$ – number of replication at each corner point
\end{itemize}

Assuming appropriate levels for each hyperparameter domain, Table \ref{tab:GSRun} presents the total number of iterations required for both RSM and GS when tuning seven hyperparameters. For GS to be effective, there must be three or more levels per hyperparameter. As a result, RSM achieves a significant reduction in the number of iterations required, ranging from $86.8\%$ to $98.2\%$ compared to GS.
\begin{table}
    \centering
    \caption{Number of Iterations of Grid Search and RSM}
    \begin{tabular}{|c|c|c|c|c|}
    \hline
        Method & GS two-level & GS three-level & GS four-level & RSM\\
        \hline
        Runs & $2^7=128$ & $3^7=2187$ & $4^7=16384$ & 288\\
        \hline
    \end{tabular}
    \label{tab:GSRun}
\end{table}

\section*{Acknowledgment}

I would like to express my sincere gratitude to my advisor for his invaluable guidance throughout the research process. I greatly appreciate your patience and support.

\vspace{12pt}
\color{red}

\end{document}